# Space-Time Localization of Inner Heliospheric Plasma Turbulence Using Multiple Spacecraft Radio Links

Adam C. Richie-Halford
Los Angeles Air Force Base, El Segundo, CA 90245

L. Iess
Dipartimento di Ingegneria Aerospaziale e Astronautica, Università di Roma "La Sapienza," I-00184 Roma, Italy

P. Tortora
DIEM -- II Facoltà di Ingegneria, Università di Bologna, I-47110 Forlì, Italy

J. W. Armstrong, S. W. Asmar, Richard Woo
Jet Propulsion Laboratory, California Institute of Technology, Pasadena CA 91109

Shadia Rifai Habbal
Institute of Astronomy, University of Hawaii

Huw Morgan
Institute of Astronomy, University of Hawaii, and Aberystwyth University, Wales, UK

## Abstract

Radio remote sensing of the heliosphere using spacecraft radio signals has been used to study the near-sun plasma in and out of the ecliptic, close to the sun, and on spatial and temporal scales not accessible with other techniques. Studies of space-time variations in the inner solar wind are particularly timely because of the desire to understand and predict space weather, which can disturb satellites and systems at 1AU and affect human space exploration. Here we demonstrate proof-of-concept of a new radio science application for spacecraft radio science links. The differing transfer functions of plasma irregularities to spacecraft radio up- and downlinks can be exploited to localize plasma scattering along the line of sight. We demonstrate the utility of this idea using Cassini radio data taken in 2001-2002. Under favorable circumstances we demonstrate how this technique, unlike other remote sensing methods, can determine center-of-scattering position to within a few thousandths of an AU and thickness of scattering region to less than about 0.02 AU. This method, applied to large data sets and used in conjunction with other solar remote sensing data such as white light data, has space weather application in studies of inhomogeneity and nonstationarity in the near-sun solar wind.

## 1. Introduction

Radio links to and from deep space probes are used for spacecraft command and control, telemetry, navigation, and radio science. Radio science applications include determination of planetary and satellite gravitational fields, measurements of properties of planetary atmospheres, ionospheres, and rings, studies of the solar wind, solar system tests of relativistic gravity, and searches for low-frequency gravitational radiation [e.g., *Tyler et al.* 2001, *Kliore et al.* 2004]. In addition to probing otherwise inaccessible regions of the solar system, the link signal-to-noise ratios (SNRs) and instrumental frequency stability are often excellent and allow sensitive propagation measurements [*Asmar et al.* 2005]. For heliospheric plasma investigations, radio propagation observations have been used to study the solar wind plasma over a wide range of distances from the sun, in- and out-of-the-ecliptic, and on spatial and temporal scales which cannot be measured with other techniques.

The high SNR and excellent time resolution of radio data lend themselves to studies of nonstationarity and inhomogeneity of the inner heliospheric plasma. Interest in space-time variations in the inner solar wind has been invigorated by the desire to understand and predict (via propagation or co-rotation) space weather, which can disturb satellites and systems at the earth, affect human space exploration, and (via intensity and phase scintillation) cause significant disruption of deep space telecommunications [*Woo* 2007 and references therein]. In this paper we present a new technique, based on processing of multiple simultaneous radio links to and from a spacecraft, for space-time localization of inner heliospheric plasma disturbances.

## 2. Transfer Functions of Plasma Irregularities to Doppler Links

At microwave frequencies solar-wind-induced refractive index variations are ($\mathbf{r}$) = $-\lambda^2 r_e \delta n_e(\mathbf{r})/(2\pi)$, where $\lambda$ is the radio wavelength, $r_e$ is the classical electron radius, and $\delta n_e$ is the electron density fluctuation at three-dimension position vector, $\mathbf{r}$. A monochromatic wave propagating through a thin scattering region ("screen") having thickness $\Delta z$ develops a geometrical optics phase shift $(2\pi/\lambda)$ ($\mathbf{r}$) $\Delta z$. This phase shift is the same for a wave propagating either "up" or "down" through the screen at point **r**. This attribute can be used for space-time localization of plasma irregularities.

Although the application to plasma localization given here is new, the idea was originally used in noise budget analysis of precision Doppler tracking experiments [*Estabrook & Wahlquist* 1975, *Estabrook* 1978, *Vessot & Levine* 1978]. The transfer function of plasma irregularities to the observed time series depends on the tracking mode. When spacecraft observations are in the two-way mode (downlink radio signal phase-locked to an uplink radio transmission) plasma fluctuations have a "two-pulse" response in the Doppler frequency time series [*Estabrook & Wahlquist* 1975, *Estabrook* 1978]. This is illustrated in Figure 1. The upper plot is a space-time diagram, showing the ground station continuously transmitting a signal to the spacecraft and continuously receiving a signal from that spacecraft. Suppose that the radio center frequencies of the up- and downlink signals are the same. In the two-way mode, the Doppler time series $y_2(t)$ is the difference between the frequency of the received downlink signal and the frequency of a ground reference oscillator. For two-way observations the ground reference oscillator also provided the signal transmitted to the spacecraft a two-way light time, $T_2$, earlier. A localized plasma blob at distance x along the line-of-sight perturbs the phase on both the up- and downlink, as illustrated, giving rise to two events in the two-way tracking time series separated by a time lag depending on the earth-blob distance: $T_2 - 2\ x/c$.

Comparison of two-way and downlink plasma time series have been used to localize a dominant plasma screen [e. g., *Armstrong* 2006]. In some tracking situations, however, more information is available. The data analyzed here were taken with the five-link Cassini radio system [*Kliore et al.* 2004]. With this radio configuration the plasma contribution to the up- and downlinks, $y_{up}(t)$ and $y_{dn}(t)$, can be computed separately [*Iess et al.* 2003, *Bertotti et al.* 2003, *Tortora et al.* 2004]. The times series $y_{up}(t)$ and $y_{dn}(t)$ respond to a localized plasma blob with one event in each time series. These events are also separated in time by $T_2 - 2\ x/c$ (lower panels in Figure 1). By cross correlating the up- and downlink Doppler time series the time separation of the plasma events can be measured and hence the plasma blob's distance from the earth determined. Since the plane-of-sky position is known (we point the ground antenna at the accurately-known spacecraft position), this technique allows localization of plasma events in time and three space dimensions.

In the idealized case of a geometrically thin screen, the uplink and downlink beams cross the thin screen at exactly one point, the phase shifts of up- and downlink are exactly the same, and the time series of up and downlink Doppler are exact, temporally offset, copies. The value of the up- and downlink Doppler cross correlation function is then unity at the correct $T_2 - 2$ x/c lag. If the screen has finite thickness, different space-time line segments contribute to the up- and downlinks. Since the line segments traversed by the up- and downlinks in the thick screen case do not exactly overlap in space-time, the up- and downlink times series are no longer offset copies of each other and peak crosscorrelation values will be less than unity. The crosscorrelation function's (ccf's) difference from unity can be used to bound the thickness of a screen dominating the observed Doppler scintillation, as discussed in the Appendix.

## 3. Observations and Signal Processing

The data used in this proof-of-concept study are from superior conjunctions tracks in 2001 and 2002 taken for the Cassini relativity experiment [*Tortora et al.* 2002, *Bertotti et al.* 2003]. During these solar conjunction intervals (+/- 15 days from conjunction during 2002) the spacecraft was several astronomical units from the earth. The observations used the full 5-link Cassini radio system [*Kliore et al.* 2000, *Bertotti et al.* 2003]: X-band downlink (≈ 8.4 GHz) coherent with X-band uplink (≈7.2 GHz), Ka-band downlink (≈32 GHz) coherent with Ka-band uplink (≈34 GHz), and a Ka-band downlink coherent with the X-band uplink. All data used here were taken with the NASA/JPL Deep Space Network 34-meter tracking antenna DSS25. The 5-link data allow plasma contributions to the Doppler, referenced here to X-band, to be separately computed for the up- and downlinks [*Tortora et al.* 2003]. In this initial study we band-pass filtered $y_{up}(t)$ and $y_{dn}(t)$ separately using arbitrary-but-reasonable band edge frequencies of 0.00056 and 0.02 Hz. This filtering, the intrinsically large SNRs of the Cassini links, and the excellent frequency stability of the ground and spacecraft systems allowed accurate estimation of the plasma contribution: the rms plasma signal is more than 300 times larger than the rms non-plasma estimation error 'noise' in the data shown here.

To implement the method we computed the cross-correlation function (ccf) between $y_{up}(t)$ and $y_{dn}(t)$ as follows. Time series $y_{dn}(t')$ was multiplied by a triangular time window, $\Lambda((t - t') / 1800$ seconds)), centered on reference downlink time, t. (Here $\Lambda(t) = 0$ if $| t | > 1$ and $1 - | t |$ if $| t | < 1$.) Similarly $y_{up}(t')$ was windowed with $\Lambda((t - t' - T_2 + 1000$ seconds)/1800 seconds). The 1000 second offset was chosen because of the *a priori* expectation that the main contribution would be near 1 AU from the earth (1 AU/speed of light ≈ 500 seconds). These windowed time series were crosscorrelated, giving the cross correlation function as a function of time lag, τ, and reference downlink time, t. Time lag was converted to line of sight range, x: $\tau = T_2 - 2\, x/c$. The triangular time windows were then advanced by 10 seconds for each of the up- and down link time series and the process repeated. This gave the plasma-contribution ccf as a function of distance along the line of sight and time throughout the tracking pass.

## 4. Examples and Discussion

Figure 2 shows the band-pass-filtered up- and downlink plasma-contribution time series for a tracking pass on 2001 May 29 (day-of-year, DOY, 149; sun-earth-spacecraft angle ≈ 6.6 degrees; $T_2 \approx 6825$ seconds, varying by ≈2 seconds over the pass.). The data show clear temporal nonstationarity, with higher variability at the start and end of the pass and lower variability in the middle of the pass. Even by eye, there is clear positive correlation between the up- and downlink time series, including several high SNR discrete 'events' (e.g. near 1730 UT, 2230 UT, and 2300 UT in the downlink, with corresponding events $\sim T_2 - 1000$ seconds later in the uplink).

Figure 3 shows the space-time correlation function for these 2001 DOY 149 data. We plot contours down to 0.4 (pilot studies cross correlating these time series with zero time lag between them suggested correlations up to about 0.3 could occur by chance.) Figure 3 has two principal features. First, there is high correlation – greater than 0.95 near 1730 UT and peaking at ≈0.99, e.g., for identifiable events near 2200-2300 UT -- over much of the track. Second, the lags where the correlation function peaks indicate plasma disturbances mostly located systematically farther away than the raypath's closest approach point to the sun at ≈1AU cos(6.6 degrees) = 0.99 AU. The regions where the correlation is high (e.g. > 0.9) give upper bounds on the effective screen thickness (see Appendix). Simulations of Kolmogorov turbulence in a uniform-thickness screen at 1 AU (and with the same signal processing used for the actual time series) give, as expected, correlation of unity when the screen is geometrically thin. Increasing the full-width screen thickness to ≈0.05 AU gives peak correlation ≈0.9. Thus, for the signal processing parameters in this pilot analysis, we estimate the full-width scattering region thickness in the high-correlation (>0.9) intervals of Figure 3 to be less than or ≈0.05 AU = 7.5 million km. During some parts of the 2001 DOY 149 track the screen thickness appears to be smaller than 0.05 AU. The region between ≈2210-2320 UT, for example, has very high correlation (greater than 0.97) and a perhaps slightly trending range (as determined from the lag of the ccf's peak) from about 1.025 AU near 2210 UT to about 1.03 AU near 2320 UT. The > 0.97 correlation suggests, from the finite-thickness screen simulations, that the full width of the region dominating the plasma scattering is less than ≈0.02 AU over this entire time interval. We also can estimate the accuracy of the fiducial range from the up- and downlink cross spectrum, under the assumption that there is a unique range-to-screen over this interval (see Appendix). For this highly-correlated interval the range is particularly well-determined: 1.028 +/- 0.003 AU.

We looked at coronal images to see if there is an obvious plane-of-sky counterpart for these disturbances on 2001 DOY 149. These disturbances could be associated with the line of sight crossing the (mainly face-on) heliospheric current sheet (HCS) in the inner corona [*Woo et al.* 1995]; the relevant scattering region may be associated with disturbances in and around the heliospheric plasma sheet (HPS) which surrounds the HCS [*Winterhalter et al.* 1994]. Figure 4 shows a map of the coronal density structure at a height of 4 solar radii (Rs) for Carrington Rotation 1976, calculated using a solar rotational tomography technique [*Morgan et al.* 2009] from ~2 weeks of the LASCO/SOHO C2 [*Brueckner et al.* 1995] coronagraph observations. Red is high density, black is low. The x-axis refers to Carrington rotation longitude (CRL), the y-axis to solar latitude. During 2001 May 29, the meridional CRL (along the Earth-Sun line) is 54 degrees. The point along the Earth-Cassini line of sight closest to the sun is at CRL 324 degrees. This point is shown as a diamond on the map. The three dotted and dashed lines are the position of the HCS, calculated using a potential field source surface (PFSS) extrapolation of photospheric magnetic field observations made by the Wilcox Solar Observatory [*Altschuler and Newkirk* 1969, *Schatten et al.* 1969, *Wang and Sheeley* 1992]. The three lines give different positions of the HCS when different boundary conditions are applied. The reasonable agreement between the PFSS and tomography results at mid-to-high latitudes show that the estimate of the HCS position is fairly accurate (PFSS is not always accurate at times outside solar minimum: see

*Morgan and Habbal* [2007]), therefore the subsolar point along the Earth-Cassini line of sight must be close to the HCS during 2001 May 29. (For various reasons, the tomography method fails near the equator, so should not be trusted there.) The thicknesses of the HCS and HPS vary by an order of magnitude as observed by spacecraft at 1 and 5AU [*Winterhalter et al.* 1994, *Smith* 2001, Zhou et al 2005]. We note that our approximate upper bound of 0.02 AU for the thickness of the 2001 DOY 149 near-sun scattering region is consistent with (about a factor of 2 larger than) the largest HPS thickness observed at 1 AU, but much larger than the ~0.001 AU *median* HPS thickness at 1AU [*Winterhalter et al.* 1994, Zhou et al. 2005].

The data in Figures 2-3 show a relatively simple case of well-localized, dominant scattering and are exemplary of the method. The technique is also diagnostic in more complicated situations. Figure 5 shows the ccf for 2002 DOY 160 (= 2002 June 9; sun-earth-spacecraft angle ≈ 9.5 degrees; $T_2 \approx 8365$ seconds.) In contrast with 2001 DOY 149, the correlation levels are lower, indicative of less-well-localized scattering on this day Especially during the middle part of this track, the correlation is comparable to or below the contouring threshold (0.4), indicating substantially extended scattering along the line-of-sight. Indeed we have data from tracks where the scintillation level, as evidenced from the rms Doppler fluctuation, is substantial but the contribution is poorly localized over an entire ~8 hour tracking pass.

Finally, the technique is still in development and our current understanding of its strengths and limitations is not complete. It is, however, clear that localization will not work well when the temporal duration of a plasma fluctuation is long compared with the earth-spacecraft light time. In this limit, the width of the Doppler disturbances on the up and downlinks overlap and signature is lost. (This is not a practical problem for space weather studies.) We have done preliminary studies of the inversion of idealized thick-screen scattering distributions. In suitable situations the square of the Fourier transform of the scattering distribution along the line of sight multiplies the up- and downlink squared coherency spectrum. Using this, one could recover a measure of an extended medium's line-of-sight spatial distribution. Finally, we have not yet experimented with pre-filtering of the data. In this pilot study we chose a reasonable passband to demonstrate the method clearly but no attempt was made at optimization.  By looking at a different region of the fluctuation spectrum, or by prewhitening the times series, we may be able to improve resolution.

## 5. Summary

This paper gave a proof-of-concept demonstration of a technique to localize inner heliospheric plasma disturbances in space and time.  The method is based on the differing transfer functions of plasma scintillation to one- and two-way radio links between the earth and a distant spacecraft.  In the technique's simplest form, discussed here, the up- and downlink plasma time series are compared to localize dominant plasma irregularities in time and along the line-of-sight.  Examples were shown for a situation where the scattering is dominated by a thin screen at well-defined location (Figure 3) and a situation where the scattering is more extended (Figure 5).  When combined with other remote sensing observations such as white light images (and other simultaneous radio observations--e.g. intensity scintillation), this method has application in studies of inhomogeneity, nonstationarity, and other manifestations of inner heliospheric plasma variability.

***Acknowledgements***:  We have greatly benefited from discussions with Frank B. Estabrook about the method.  We thank W. A. Coles, B. J. Rickett, S. R. Spangler, and an anonymous referee for comments on the paper.  ARH's work was performed at the Jet Propulsion Laboratory while on a US Air Force-JPL employee exchange program.  LI and PT were supported in part by the Italian Space Agency (ASI).  SH's and HM's research was supported by NASA grant NNX07AH90G to the University of Hawaii.  For JWA, SWA and RW the research described here was carried out at the Jet Propulsion Laboratory, California Institute of Technology, under a contract with the National Aeronautics and Space Administration. We also thank the personnel of the Cassini Project, the JPL Radio Science Systems Group, and the NASA/JPL Deep Space Network. The SOHO/LASCO data used here are produced by a consortium of the Naval Research Laboratory (USA), Max-Planck-Institut fur Aeronomie (Germany), Laboratoire d'Astronomie (France), and the University of Birmingham (UK).  SOHO is a project of international cooperation between ESA and NASA. The Wilcox Solar Observatory's photospheric field data and extrapolations were obtained from the WSO section of Stanford University's website courtesy of J. T. Hoeksema. WSO is supported by NASA, the NSF, and ONR.

**Appendix**: Distance Uncertainty and Screen Thickness Estimate

The accuracy of the ccf time lag, hence the accuracy of the range to a plasma screen, can be estimated from the cross spectrum of the up- and downlink time series. Assume there is a thin screen (justified empirically in cases where the peak correlation of the ccf is close to unity), hence a unique distance of the plasma disturbances from earth. Then there is a unique time lag in the ccf associated with that screen: the up- and downlink time series are copies of themselves offset by $\tau_{true} = T_2 - 2x/c$. In real situations approximating this idealization (e.g. the data between ≈2210-2320 on 2001 DOY 149), the lag will be nonunique due to variability of the true range over the time interval, finite thickness of the "screen", and estimation error of the ccf. This nonuniqueness reflects itself in range uncertainty.

We can estimate range accuracy from the statistics of the cross-spectral phase estimates [*Jenkins and Watts* 1969]. The idea is to pre-align the two time series, $y_{up}$ and $y_{dn}$ by an initial estimate of the true offset, $\tau^*$, determined for example from the peak of the average ccf over the time interval. This initial lag estimate will in general have an error: $\tau^* = \tau_{true} + \delta\tau$. If $\delta\tau$ is non-zero then, from the shift theorem for Fourier transforms [*Bracewell* 1965], there will be a non-zero slope in the cross spectral phase (=inverse tangent of Im[C(f)]/Re[C(f)], where C(f) is the cross-spectrum). The numerical value of the slope will be $2\pi\,\delta\tau$, from which the correction $\delta\tau$ can be estimated. The uncertainty in the final best-fit lag can then be determined from the uncertainty in the slope of the cross-spectral phase.

Consider the downlink data during the received time interval 2210-2320 UT and its uplink counterpart in the interval ~$T_2$-1000 seconds later. Based on the ccf in this time interval, the uplink time series was pre-aligned relative to the downlink by $\tau^* = 5796$ seconds. The smoothed auto spectra and smoothed cross spectrum were estimated in three separate procedures by Fourier transforming blocks of data 1024, 512, and 256 seconds long, giving 4+, 8+, and 16+ averages during the 2210-2320 UT time interval. Rectangular transform windows were used, giving frequency resolution of 1/1024sec, 1/512sec, and 1/256 sec, respectively. Smoothing was done via simple averages of the Fourier transform-squared (for the auto spectra) and separate averages of the real and imaginary parts of the cross spectrum. The smoothed squared-coherency, $\kappa_{12}^2(f)$, was estimated by forming the modulus-squared of the cross spectrum divided by the product of the auto spectra [*Jenkins and Watts* 1969]. The smoothed cross-spectral phase, $\phi_{12}(f)$, was computed from the inverse tangent of the ratio of the smoothed imaginary part of the cross spectrum to the smoothed real part. Figure A1 shows these quantities for the three frequency resolutions. The slope of the cross-spectral phase in the Fourier band 0-0.01 Hz, after alignment by 5796 seconds, is non-zero and indicates an additional offset of 0.1 sec would be required for best alignment. The formal standard deviation in the slope of the cross-spectral phase is small (corresponding to a standard deviation in the estimated best-lag of less than 1 second). The error in the slope is poorly estimated, however, due in part to the small number of points going into its determination. Instead of using the formal standard deviation in the slope, we adopt a more conservative viewpoint: the uncertainty in the slope of the residual cross spectral phase in Figure A1 appears bounded by +/- 10 degrees/0.01 Hz. This gives an uncertainty in the 5796.1 second time lag estimate of < 2.8 seconds and hence a formal uncertainty in the distance (for this high-correlation interval, not for a typical interval) of less than 0.003 AU.

A finite-thickness screen will not have unity cross-correlation because the up- and downlink paths do not traverse exactly the same points in space-time (Figure 1) and thus the two time series are not simply offset copies of each other. If the screen thickness is small, the peak correlation will be only slightly smaller than unity (i.e., the practical case for the 2001 DOY 149 data). The decorrelation from unity can be used to estimate screen thickness. We simulated Kolmogorov turbulence [*Woo & Armstrong* 1979] in uniformly-weighted screens of varying thickness taking into account the differences in the ray paths of the up- and downlinks to produce synthetic Doppler time series. The simulated up- and downlink were then processed through the same software used to analyze the real data and the average peak correlation was determined over a simulated "tracking pass" of ≈50000 seconds (a little longer than a typical real tracking pass). The results are shown in Figure A2. (In the simulations, layers were added in 0.001 AU steps to the previous screen, so the points in Figure A2 are not statistically independent of each other.) The interval ≈2210-2320 on 2001 DOY 149 has peak correlation > 0.97. Although this simulation approach is model-dependent (Kolmogorov turbulence, uniformly weighted screen), comparison with Figure A2 suggests the real data arise from a region having full-width thickness ≈0.02 AU or smaller.

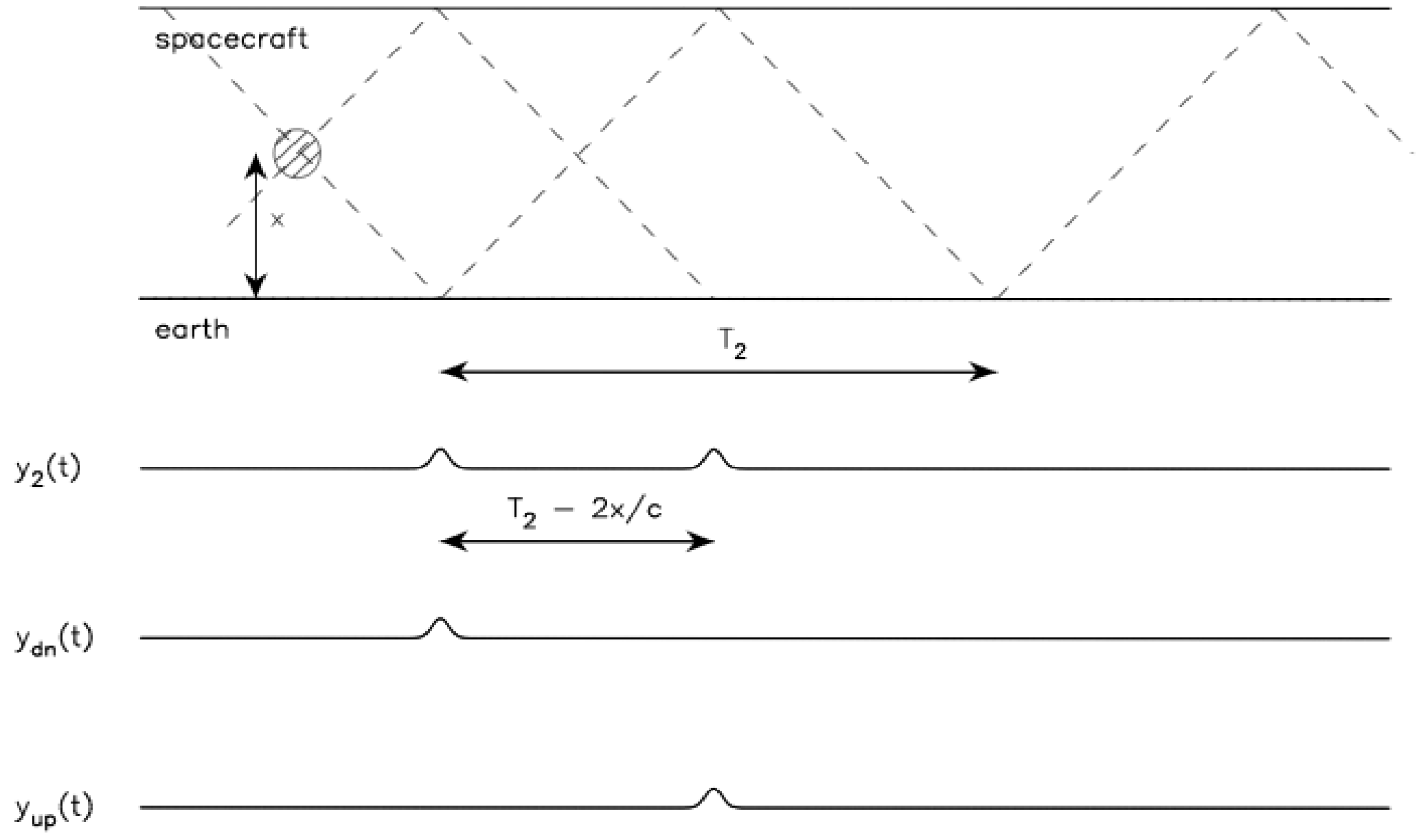

**Figure 1**. Transfer function of plasma fluctuations to one- and two-way phase/Doppler scintillation allows localization of plasma blobs along the line-of-sight. Upper plot is a space-time diagram (space vertically, time horizontally). The ground station and the spacecraft are continuously exchanging microwave signals, some of which are shown as dashed lines. If the signals pass through a well-localized plasma blob (indicated by the cross-hatched circular area) the phase is perturbed. This perturbation is observed on both the uplink and downlink signals and on the "two-way" (coherently transponded) Doppler. The two-way, $y_2(t)$, and the one-way plasma-contribution Doppler time series, $y_{up}(t)$ and $y_{dn}(t)$, are shown in the lower plot. The phase perturbation is seen initially on the downlink, and – in the two-way Doppler -- latter when the perturbation on the uplink is phase-coherently re-transmitted back to the ground. The effect in $y_2(t)$ is two positively-correlated features in the time series, separated in time by $T_2 – 2x/c$, where $T_2$ is the two-way light time and x is the distance of the blob from the earth. The one-way up and down Doppler time series detect the blob once each, but also separated by $T_2 – 2x/c$. Crosscorrelation or windowed matched filtering between pairs of Doppler time series allows estimates of the time delay, localizing the blob.

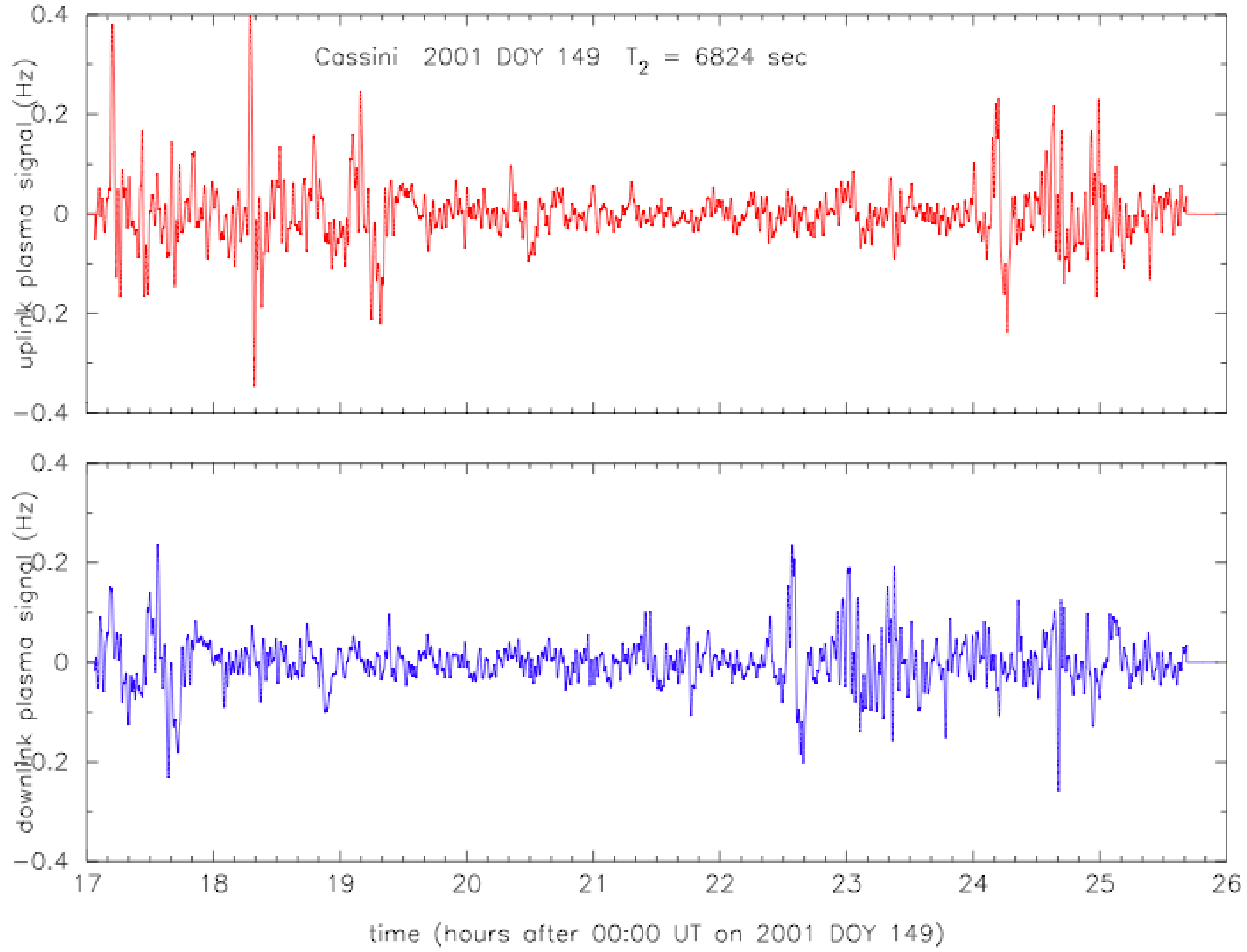

**Figure 2**. Time series of the plasma contribution to the X-band up- and down links, $y_{up}(t)$ and $y_{dn}(t)$, for DSS25 Cassini track on 2001 DOY 149 = 2001 May 29. The two-way light time $T_2$ was ≈6824 seconds. Several large amplitude features in the downlink time series (at e.g. about 1730 UT, 2230 UT, 2300 UT) have clear "echoes" in the uplink time series with time difference $\sim T_2 - 2*1.028$ AU/c. Figure 3 shows the windowed cross correlation (see text) of these two time series, quantifying the space-time location of the plasma causing the Doppler variability.

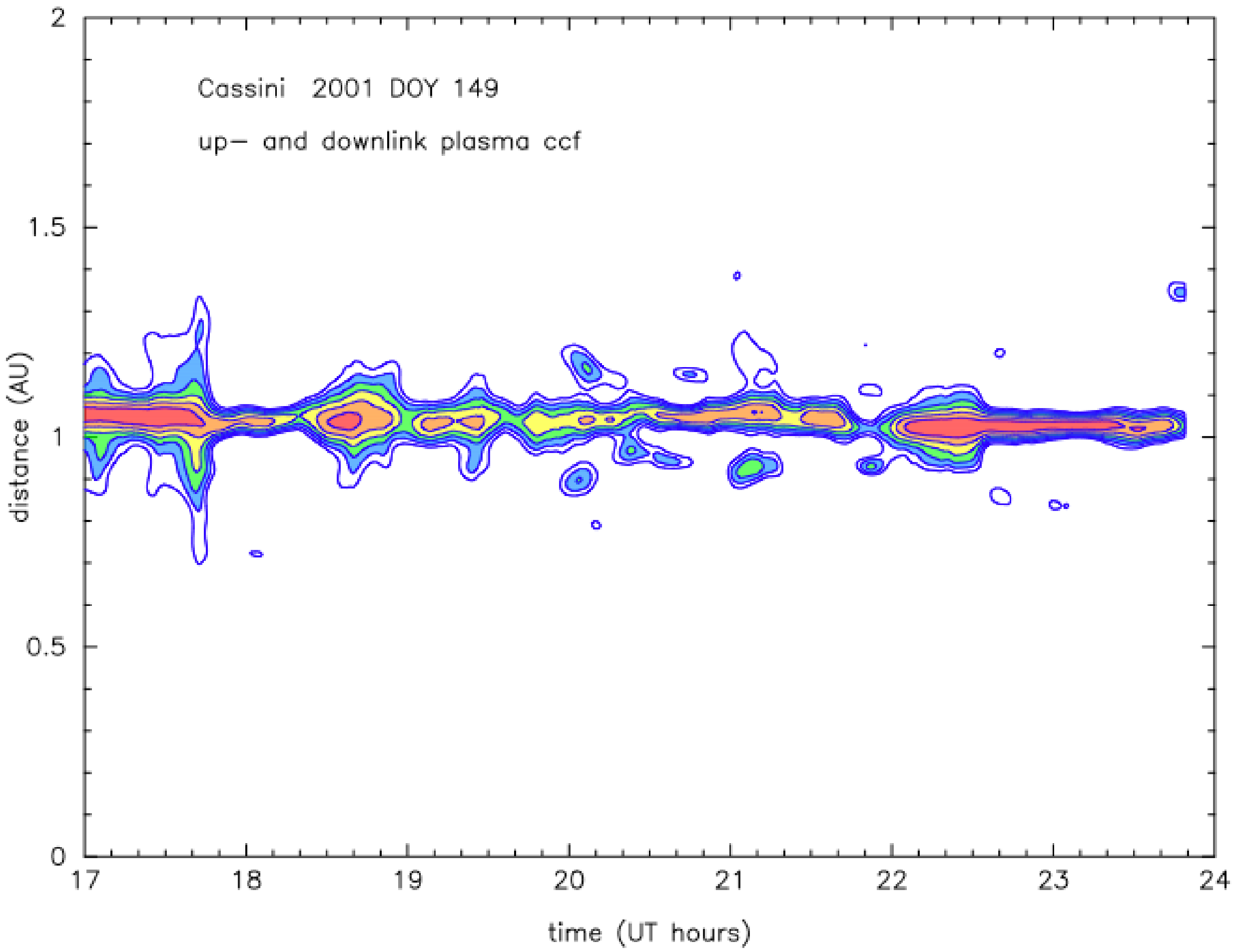

**Figure 3.** Space-time cross correlation function of the 2001 DOY 149 plasma up- and downlink time series plotted in Figure 2. Y-axis is the distance from the Earth; x-axis is downlink received time. Contours of cross correlation value are plotted between 0.9 and 0.4, in 0.1 increments. Correlations > 0.9 are shaded red, those between 0.8 and 0.9 in orange, and so forth. The accuracy of the range determination and bounds to the thickness of the region contributing the scintillation are discussed in the text.

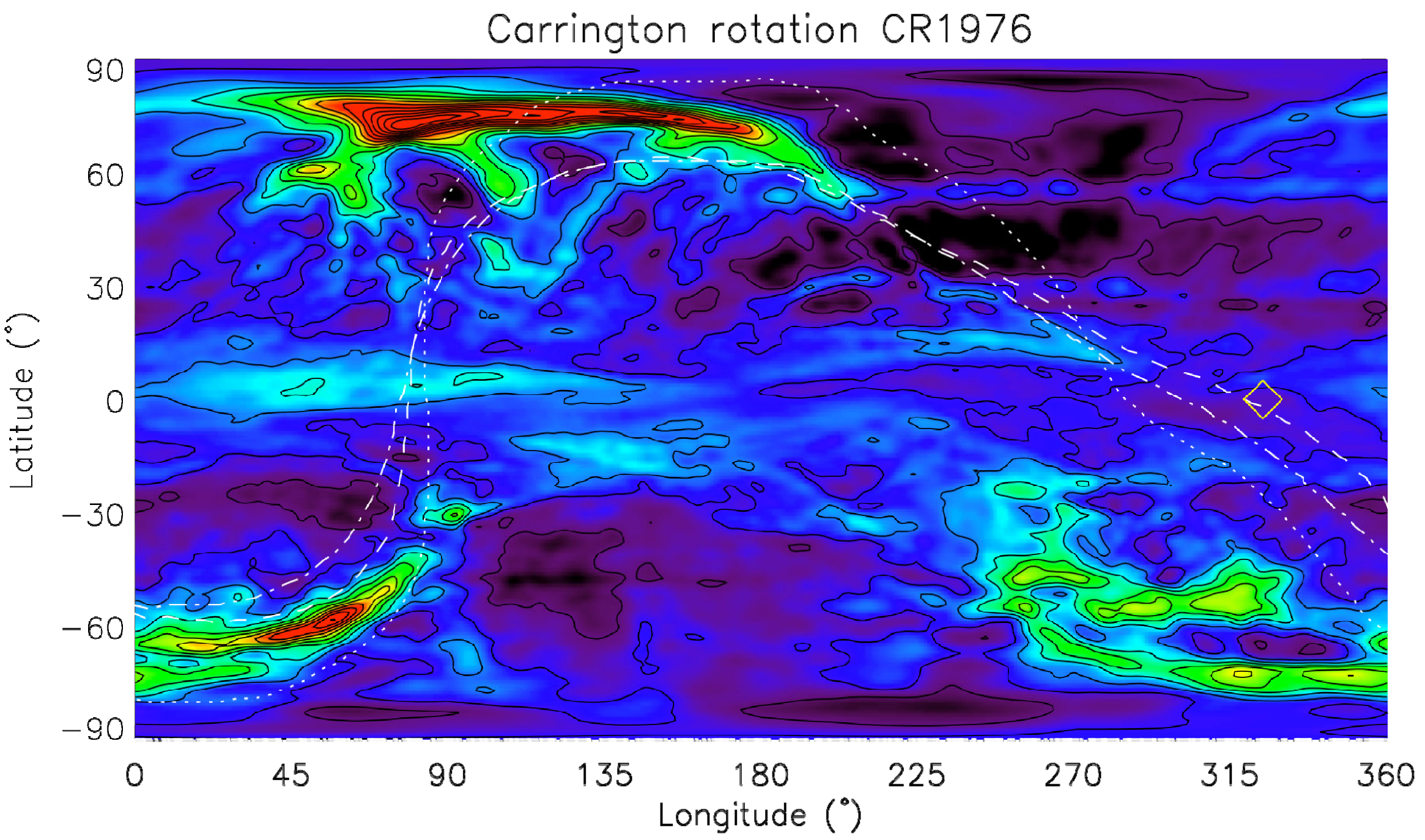

**Figure 4**. Map of coronal density structure at a height of 4 solar radii for Carrington Rotation 1976, calculated using a solar rotational tomography technique from ~2 weeks of LASCO/SOHO C2 coronagraph observations. Red is high density, black is low. The x-axis refers to Carrington longitude (CRL), the y-axis to solar latitude. The point along the earth-Cassini line of sight closest to the sun is shown as a diamond at CRL 324 degrees on the map. The three dotted and dashed lines are the position of the heliospheric current sheet (HCS), calculated using a potential field source surface (PFSS) extrapolation of photospheric magnetic field observations made by the Wilcox Solar Observatory. (The three lines give different positions of the HCS when different boundary conditions are applied. See main text.)

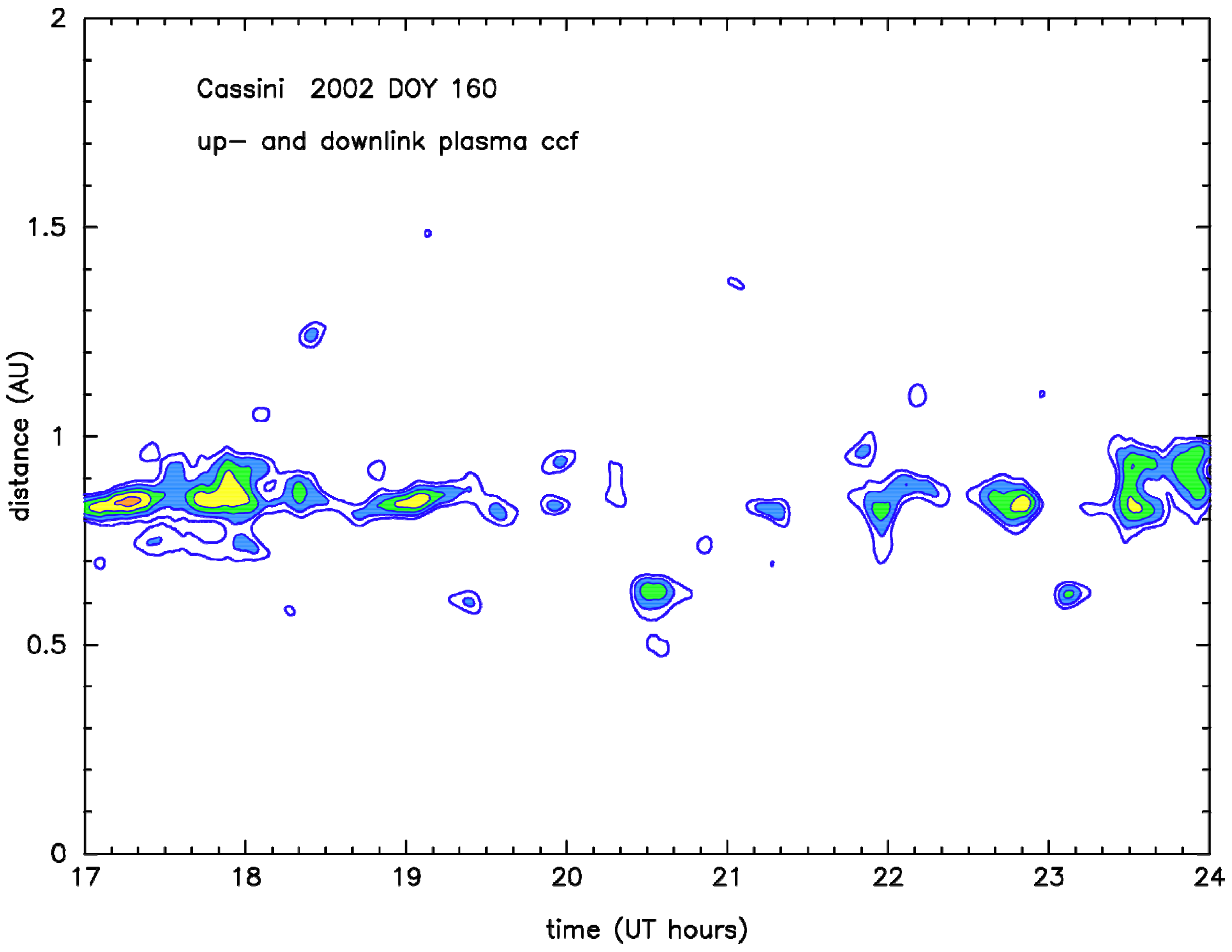

**Figure 5**. As Figure 3, but for 2002 DOY 160 = 2002 June 9. The sun-earth-spacecraft angle was about 9.5 degrees and the two-way light time was about 8365 seconds. Correlation levels in general lower than those in Figure 3. At the start of the track a scattering region nearer to the earth than the raypath closest approach point (≈ 1AU cos(9.5 degrees) = 0.986 AU) is indicated. In mid-track the correlation is particularly low, indicative of extended, rather than localized, scattering.

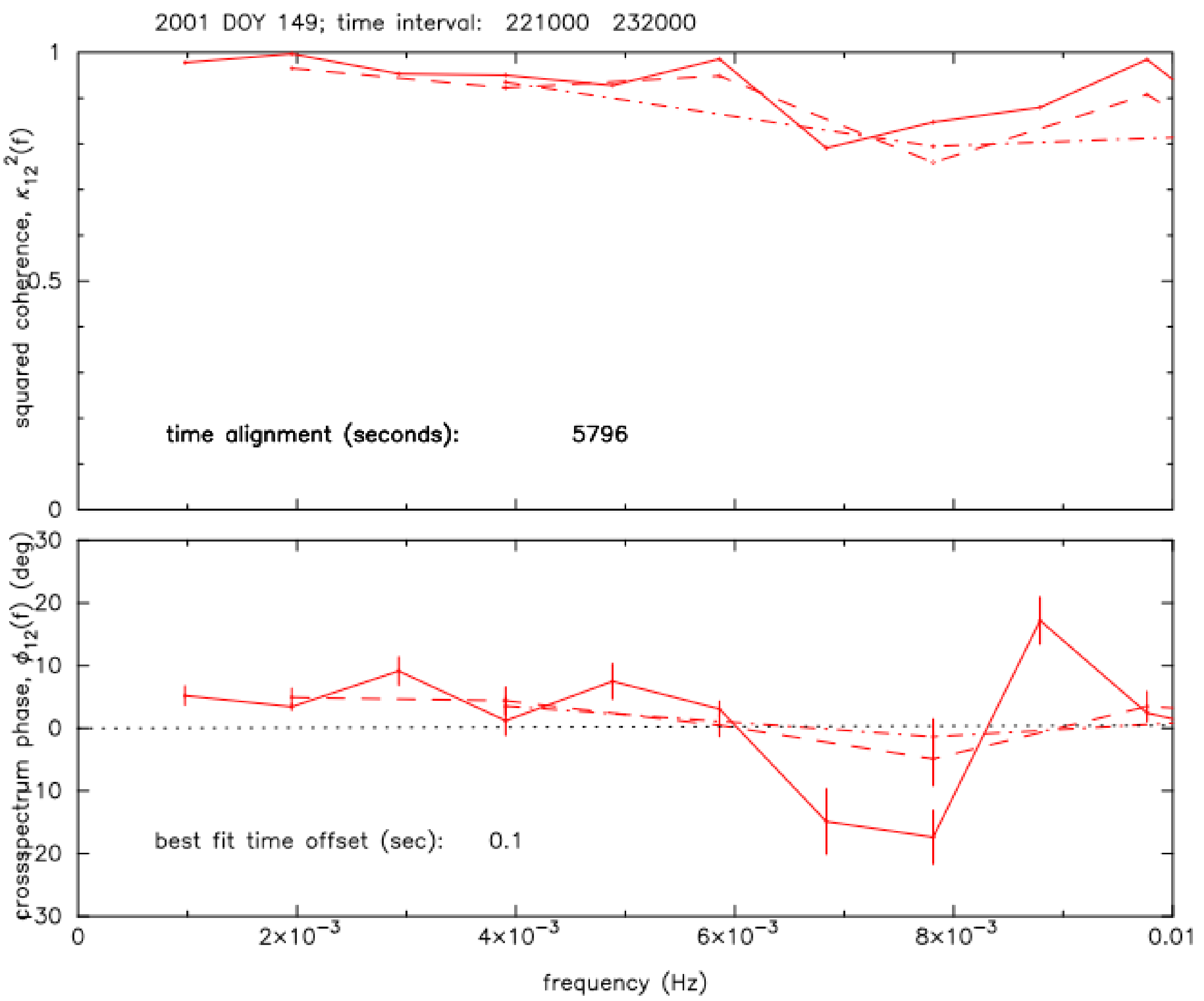

**Figure A1**. Smoothed squared coherence and smoothed cross-spectral phase for the up- and downlink plasma contributions for time interval 2001 DOY 149 2210-2320 UT. The data have been pre-aligned by 5796 seconds and analyzed with resolution bandwidths of 1/1024 Hz (solid line), 1/512 Hz (dashed line), and 1/256 Hz (dot-dash lines). Uncertainties shown for the cross spectral phase are formal +/- 3 sigma, based on the numbers of degrees of freedom in the estimates and on the squared-coherency at each frequency [*Jenkins and Watts* 1969, equation 9.2.21]. The non-zero slope in the phase spectrum gives a refined estimate for the lag giving best correlation between the up- and downlink plasma time series: 5796.1 seconds. The uncertainty in the slope gives an uncertainty in this time offset and hence an uncertainty in the distance to the screen (see Appendix).

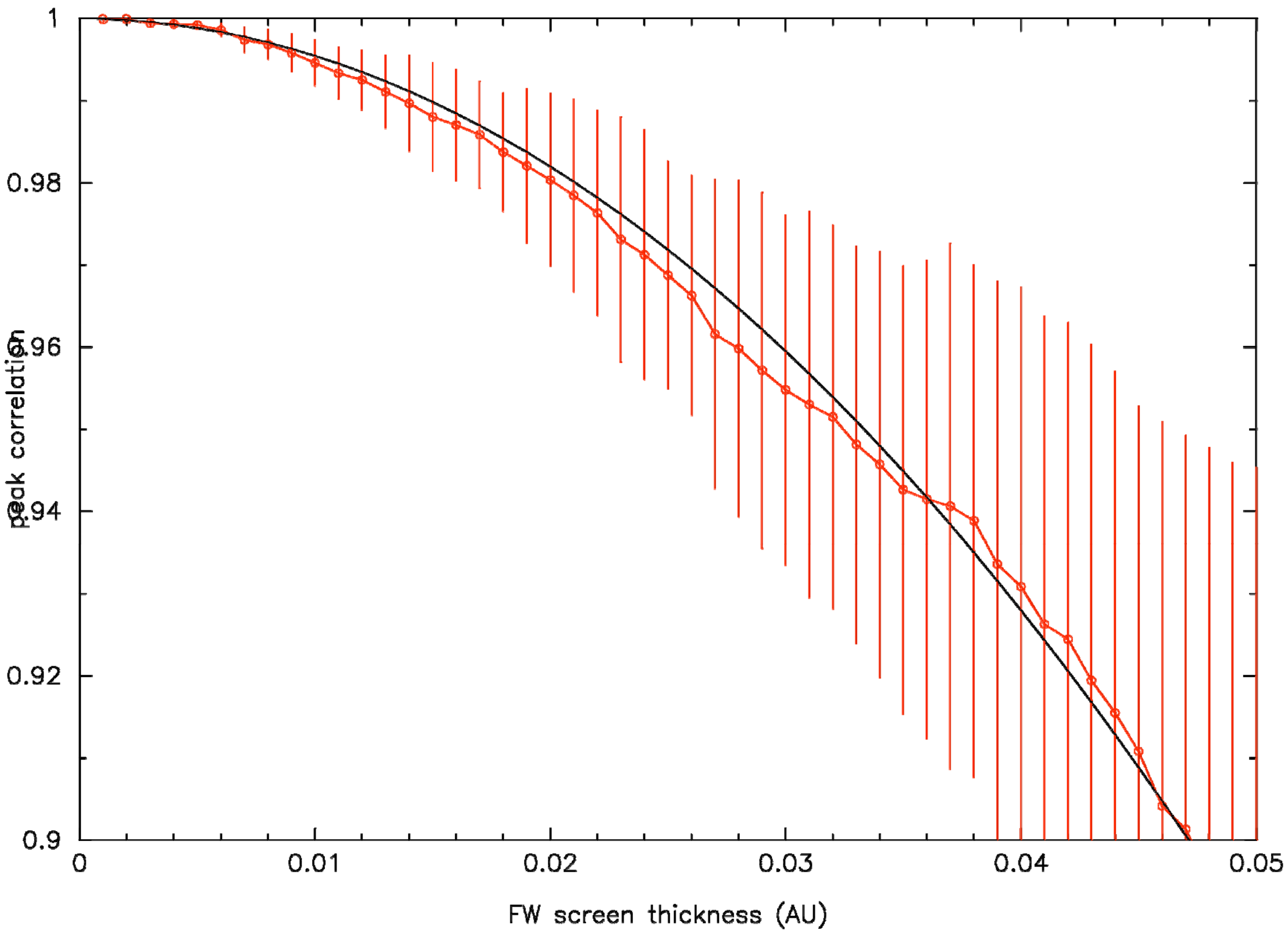

**Figure A2**. Peak correlation value versus full-width screen thickness from simulations of Kolmogorov-spectrum turbulence. Simulated data were produced from uniformly weighted layers over the indicated full-width thickness with the ”near” edge of the turbulence in the simulation at 1 AU. Red points are peak correlation and standard deviation for simulations of length slightly longer than one actual tracking pass, processed though the same software used in the analysis of the Figure 3 and 5 data. Black line is a smooth curve, not a fit. For screen thickness small compared with the time constant of the low pass filtering, cross-correlation function width is set by the low-pass filtering and its decorrelation from unity can be used to estimate screen thickness.

## *References*

Altschuler, M. D. & Newkirk, G. (1969) "Magnetic Fields and the Structure of the Solar Corona. I: Methods of Calculating Coronal Fields" *Solar Physics*, **9**, 131-149 10.1007/BF00145734

Armstrong, J.W. (2006) "Low-Frequency Gravitational Wave Searches Using Spacecraft Doppler Tracking", *Living Rev. Relativity*, **9**, 1 [Online Article]: cited 4 May 2009, http://www.livingreviews.org/lrr-2006-1

Asmar, S.W., Armstrong, J.W., Iess, L., & Tortora, P. (2005) "Spacecraft Doppler Tracking: Noise Budget and Achievable Accuracy in Precision Radio Science Observations", *Radio Science* **40**, RS2001 10.1029/2004RS003101

Bertotti, B., Iess, L., & Tortora, P. (2003) "A Test of General Relativity Using Radio Links with the Cassini Spacecraft", *Nature*, **425**, 374-376.

Bracewell, R. (1965) *The Fourier Transform and Its Applications* (McGraw-Hill: New York).

Brueckner, G. E., Howard, R. A., Koomen, M. J., Korendyke, C. M., Michels, D. J., Moses, J. D., Socker, G. G., Dere, K. P., Lamy, P. L., Llebaria, A., Bout, M. V., Schwenn, R., Simnett, G. M. Benford, D. K., & Eyles, C. J. (1995) "The Large Angle Spectroscopic Coronagraph (LASCO)", *Solar Physics*, **162**, 357-402.

Estabrook, F.B. (1978) "Gravitational Wave Detection With the Solar Probe. II. The Doppler Tracking Method" in *A Close-Up of the Sun* (JPL Publication 78-70), September 1, 1978, pp. 441-449.

Estabrook, F.B. & Wahlquist, H.D. (1975) "Response of Doppler Spacecraft Tracking to Gravitational Radiation" *Gen. Rel. Grav.*, **6**, 439-447.

Iess, L., Anderson, J. D., Asmar, S. W., Barbinis, E., Bertotti, B., Fleischman, D. U., Gatti, M. S., Goltz, G. L., Herrera, R. G., Lau, E. & Oudrhiri, K. (2003) "The Cassini Solar Conjunctions Experiment: A New Test of General Relativity", in *Proceedings of the IEEE Aerospace Conference 2003*, Big Sky, MT, March 8-15, 2003, IEEE Conference Proceedings, 1-211-223 Institute of Electrical and Electronic Engineers, Piscataway, USA, 2003)

Jenkins, G.M. & Watts, D.G. (1969) *Spectral Analysis and Its Applications* (Holden-Day: San Francisco).

Kliore, A.J., Anderson, J. D., Armstrong, J. W., Asmar, S. W., Hamilton, C. L., Rappaport, N. J., Wahlquist, H. D., Anbrosini, R., Flasar, F. M., French, R. G., Iess, L., Marouf, E. A. & Nagy, A. F. (2004) "Cassini Radio Science", *Space Science Reviews* **115**, 1-70.

Morgan, H. & Habbal, S. R. (2007) “An Empirical 3D Model of the Large-Scale Coronal Structure Based on the Distribution of $H\alpha$ Filaments on the Solar Disk”, *Astron. Astrophys*, **464**, 357-365 10.1051/0004-6361:20066482

Morgan, H., Habbal, S. R., & Lugaz, N. (2009) “Mapping the Structure of the Corona Using Fourier Backprojection Tomography”, *Ap. J.*, **690**, 1119-1129 10.1088/0004-637X/690/2/1119

Shatten, K. H., Wilcox, J. M., & Ness, N. F. (1969) “A Model of Interplanetary and Coronal Magnetic Fields”, *Solar Physics*, **6**, 442-455 10.1007/BF00146478

Smith, E. (2001) “The Heliospheric Current Sheet”, *JGR,* **106** (A8), 15819-15831

Tortora, P., Iess, L. , & Ekelund, J. E. *“Accurate Navigation of Deep Space Probes using Multifrequency Links: the Cassini Breakthrough During Solar Conjunction Experiments”*, The World Space Congress, October 10-19, 2002, Houston, USA

Tortora, P., Iess, L., & Herrera, R. G. *“The Cassini Multifrequency Link Performance During 2002 Solar Conjunction”, Proceedings of the IEEE Aerospace Conference*, March 8-15, 2003, Big Sky, Montana, USA, Vol. 3, 1465-1473

Tortora, P., Iess, L., Bordi, J.J., Ekelund, J.E., & Roth, D.C. (2004) “Precise Cassini Navigation During Solar Conjunctions through Multifrequency Plasma Calibrations”, *Journal of Guidance, Control and Dynamics*, **27**, 251-257.

Tyler, G. L., G. Balmino, D. P. Hinson, W. L. Sjogren, D. E. Smith, R. A. Simpson, S. W. Asmar, P. Priest, & J. D. Twicken (2001) “Radio Science Observations with Mars Global Surveyor: Orbit Insertion Through One Mars Year in Mapping Orbit,” *Journal of Geophysical Research - Planets* 106(E10), 23327-23348.

Vessot, R.F.C. & M.W. Levine (1978) “A Time-Correlated Four-Link Doppler Tracking System” in *A Close-Up of the Sun* (JPL Publication 78-70), September 1, 1978, pp. 457-497.

Wang, Y.-M. & Sheeley, N. R. (1992) “On Potential Field Models of the Solar Corona” *Ap. J.,* **392**, 310-319 10.1086/171430

Winterhalter, D., Smith, E. J., Burton, M. E., Murphy, N., & McComas, D. J. (1994) “The Heliospheric Plasma Sheet” *JGR*, **99** (A4), 6667-6680

Woo, R. (2007) “Space Weather and Deep Space Communications” *Space Weather*, **5**, S09004, doi:10.1029/2006SW000307

Woo, R. & Armstrong, J., W. (1979) "Spacecraft Radio Scattering Observations of the Power Spectrum of Electron Density Fluctuations in the Solar Wind", *JGR*, **84**, 7288-7296.

Woo, Richard, Armstrong J.W., & Gazis, P. (1995) “Doppler Scintillation Measurements of the Heliospheric Current Sheet and Coronal Streamers Close to the Sun” *Space Science Reviews*, **72**, 223-228.

Zhou, X.-Y., Smith, E. J., Winterhalter, D., McComas, D. J., Skoug, R. M., Goldstein, B. E. & Smith, C. W. (2005) “Morphology and Evolution of the Heliospheric Current and Plasma Sheets From 1 to 5 AU” in *Proc. Solar Wind 11-SOHO 16 “Connecting Sun and Heliosphere”*, Whistler, Canada 12-17 June 2005 (ESA SP-592, September 2005).

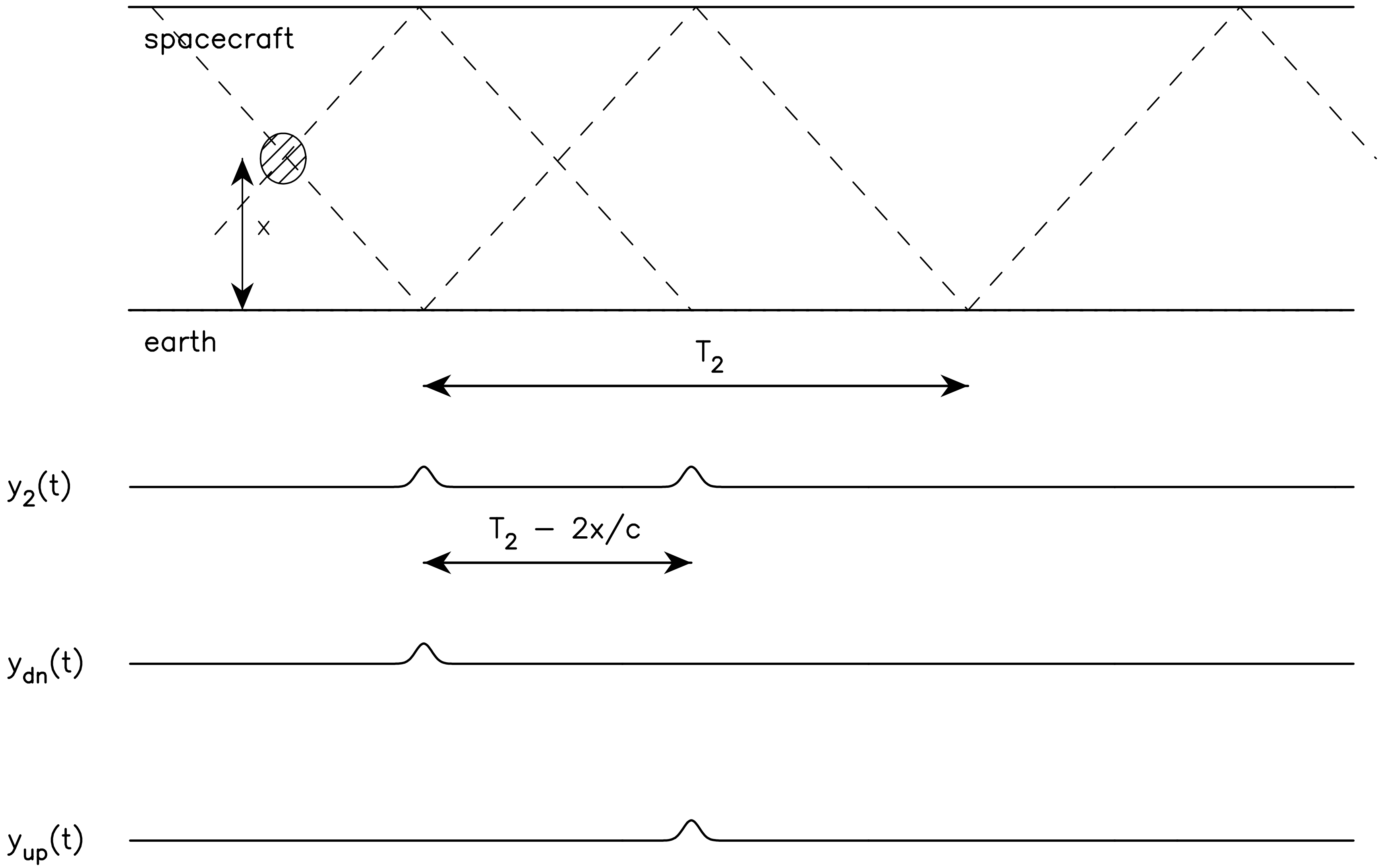

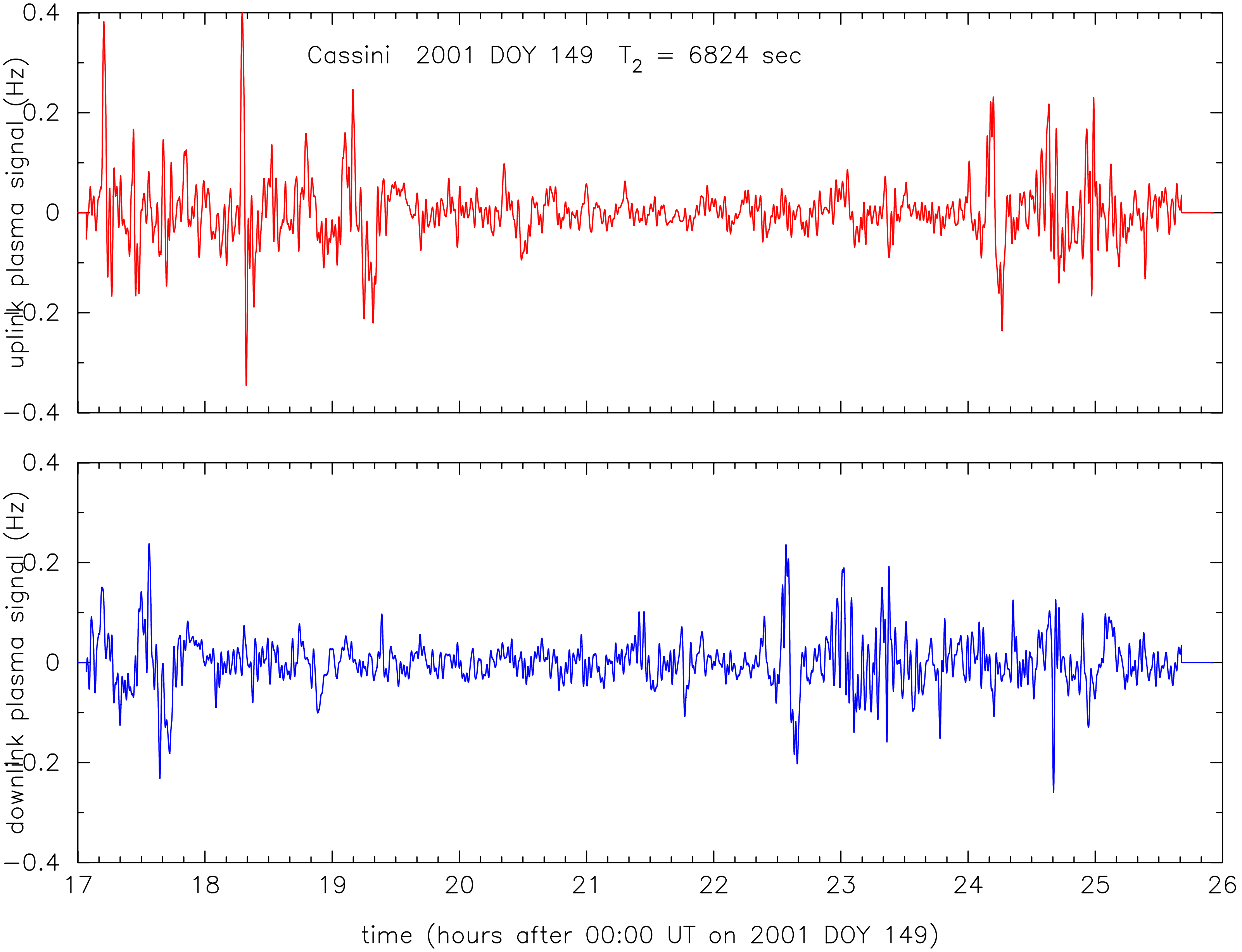

Cassini 2001 DOY 149 $T_2$ = 6824 sec
uplink plasma signal (Hz)
downlink plasma signal (Hz)
0.4
0.2
0
−0.2
−0.4
17
18
19
20
21
22
23
24
25
26
time (hours after 00:00 UT on 2001 DOY 149)

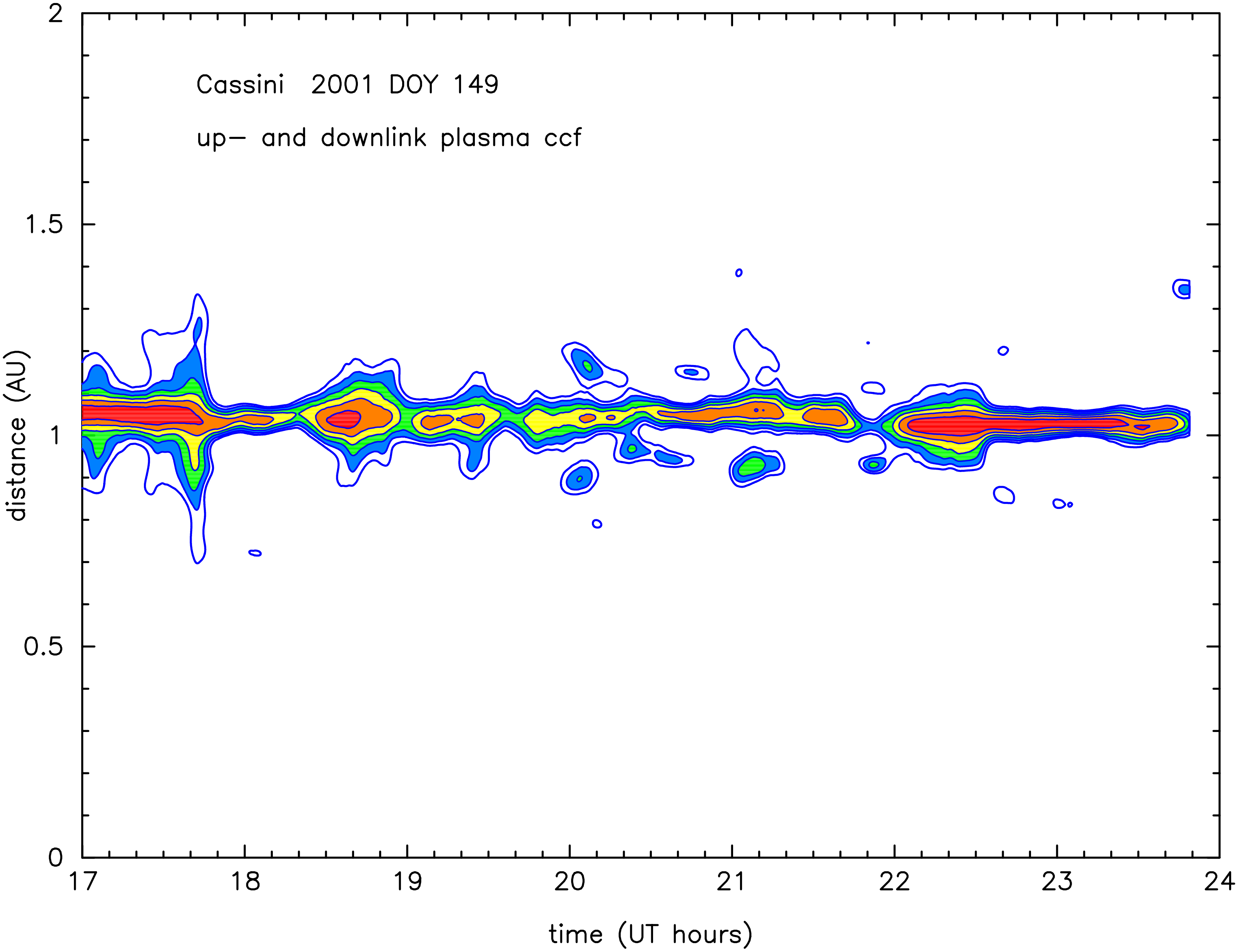

Cassini 2001 DOY 149
up– and downlink plasma ccf
distance (AU)
time (UT hours)
0
0.5
1
1.5
2
17
18
19
20
21
22
23
24

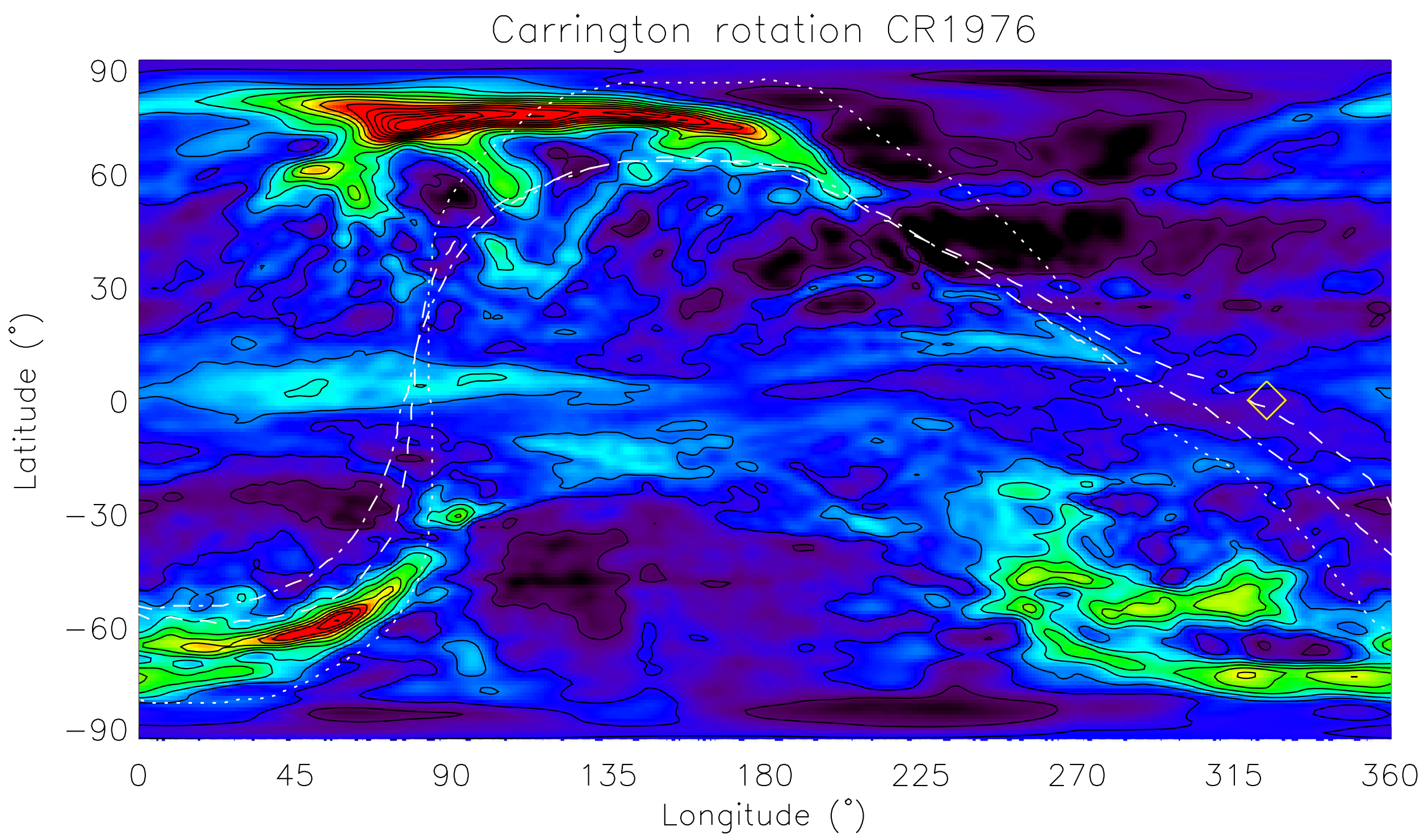
Carrington rotation CR1976
90
60
30
0
−30
−60
−90
Latitude (°)
0
45
90
135
180
225
270
315
360
Longitude (°)

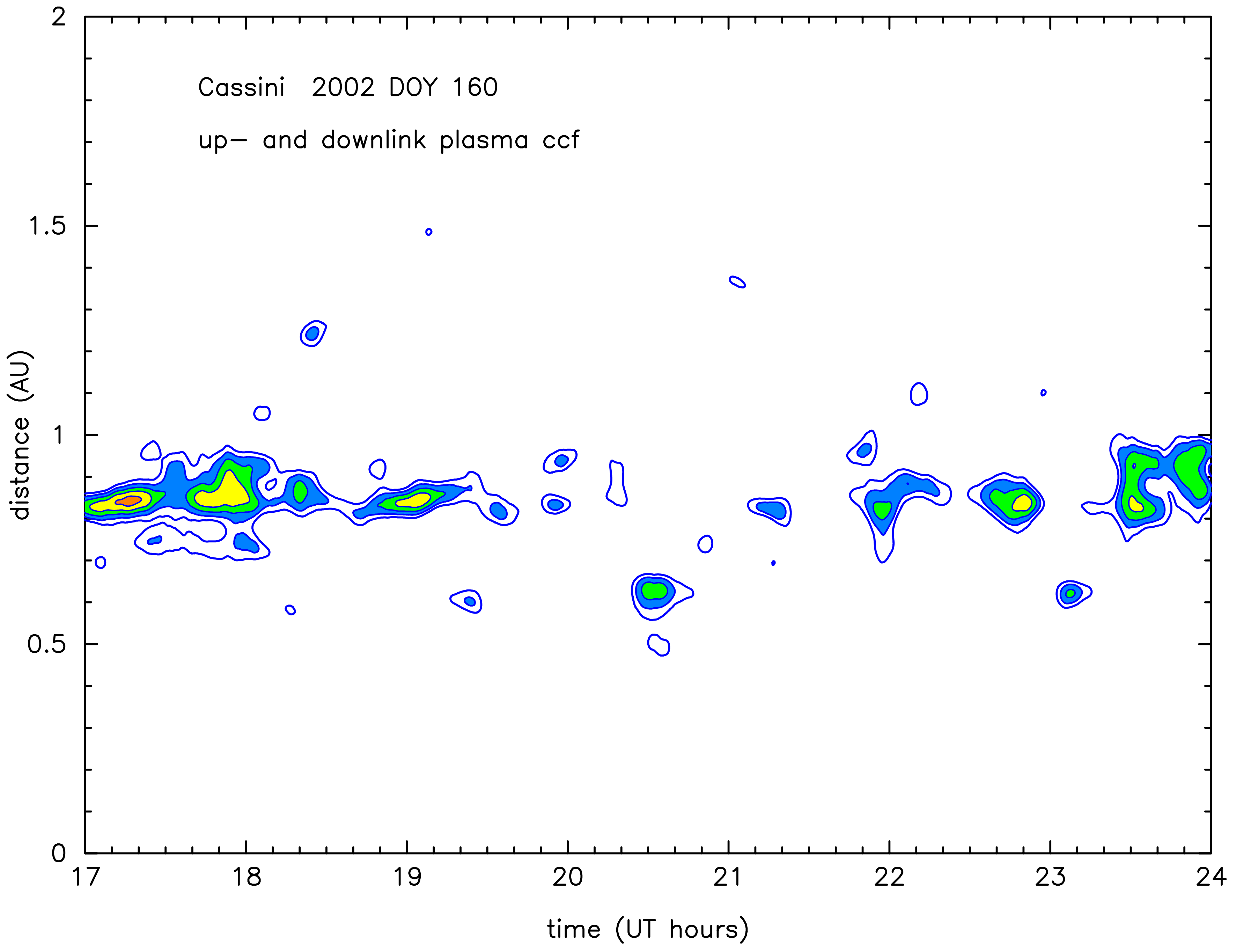

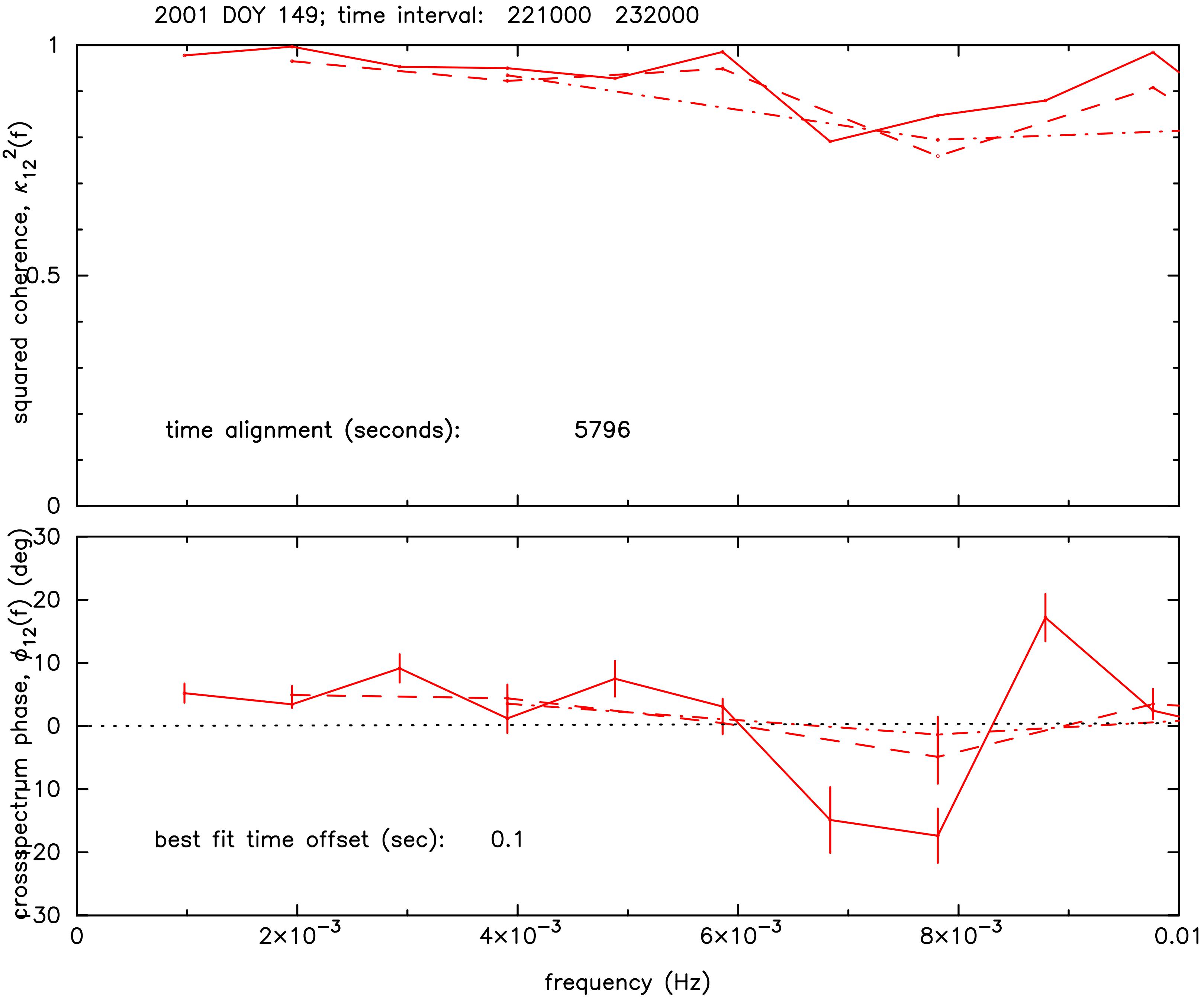

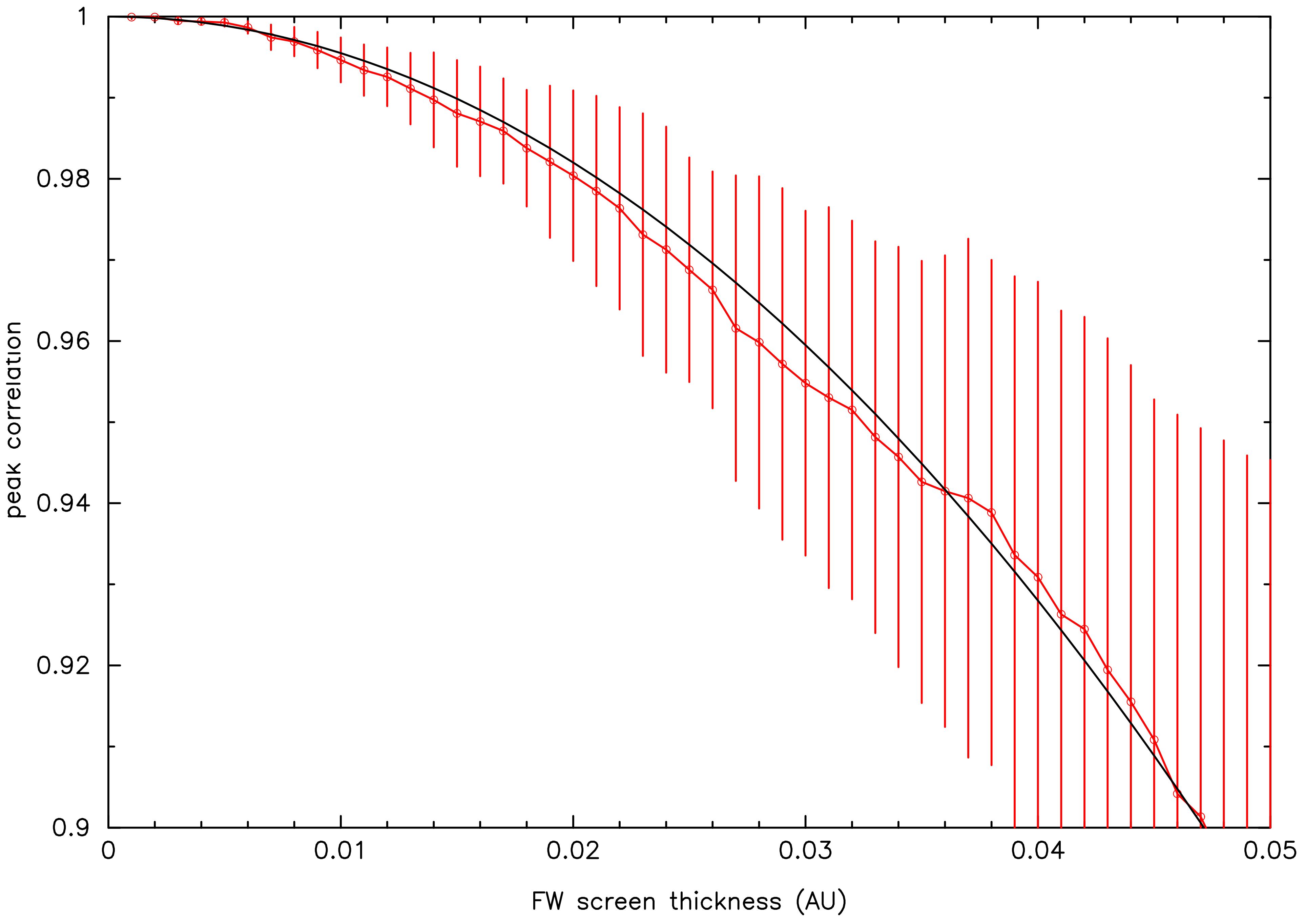